# Exciton diffusion in two-dimensional metal-halide perovskites

Michael Seitz[1,2], Alvaro J. Magdaleno[1,2], Nerea Alcázar-Cano[1,3], Marc Meléndez[1,3], Tim J. Lubbers[1,2], Sanne W. Walraven[1,2], Sahar Pakdel[4], Elsa Prada[1,2], Rafael Delgado-Buscalioni[1,3], and Ferry Prins[1,2]

*1. Condensed Matter Physics Center (IFIMAC), Autonomous University of Madrid, 28049 Madrid, Spain*

*2. Department of Condensed Matter Physics, Autonomous University of Madrid, 28049 Madrid, Spain*

*3. Department of Theoretical Condensed Matter Physics, Autonomous University of Madrid, 28049 Madrid, Spain*

*4. Department of Physics and Astronomy, Aarhus University, 8000 Aarhus C, Denmark*

*To whom correspondence should be addressed: ferry.prins@uam.es*

# Abstract

Two-dimensional perovskites, in which inorganic layers are stabilized by organic spacer molecules, are attracting increasing attention as a more robust analogue to the conventional three-dimensional metal-halide perovskites. However, reducing the perovskite dimensionality alters their optoelectronic properties dramatically, yielding excited states that are dominated by bound electron-hole pairs known as excitons, rather than by free charge carriers common to their bulk counterparts. Despite the growing interest in two-dimensional perovskites for both light harvesting and light emitting applications, the full impact of the excitonic nature on their optoelectronic properties remains unclear, particularly regarding the spatial dynamics of the excitons within the two-dimensional (2D) plane. Here, we present direct measurements of in-plane exciton transport in single-crystalline layered perovskites. Using time-resolved fluorescence microscopy, we show that excitons undergo an initial fast, intrinsic normal diffusion through the crystalline plane, followed by a transition to a slower subdiffusive regime as excitons get trapped. Interestingly, the early intrinsic exciton diffusivity depends sensitively on the choice of organic spacer. We find a clear correlation between the stiffness of the lattice and the diffusivity, suggesting exciton-phonon interactions to be dominant in determining the spatial dynamics of the excitons in these materials. Our findings provide a clear design strategy to optimize exciton transport in these systems.

# Introduction

Metal-halide perovskites are a versatile material platform for light harvesting[1–5] and light emitting applications,[6,7] combining the advantages of solution processability with high ambipolar charge carrier mobilities,[8,9] high defect tolerance,[10–12] and tunable optical properties.[13–15] Currently, the main challenge in the applicability of perovskites is their poor environmental stability.[16–19] Reducing the dimensionality of the perovskite has proven to be one of the most promising strategies to yield a more stable performance.[20–23] Perovskite solar cells with mixed 2D and 3D phases, for example, have been fabricated with >22% efficiencies[24] and stable performance for more than 10.000 hours,[25] while phase pure 2D perovskite solar cells have been reported with efficiencies above 18%.[26,27] Likewise, significant stability improvements have been reported for phase pure 2D perovskites as the light emitting layer in LED technologies.[28–34] The improved environmental stability in 2D perovskite phases is attributed to a better moisture resistance due to the hydrophobic organic spacers that passivate the inorganic perovskite sheets, as well as an increased formation energy of the material.[20–23]

However, the reduced dimensionality of 2D perovskites dramatically affects the charge carrier dynamics in the material, requiring careful consideration in their application in optoelectronic devices.[35–37] 2D perovskites are composed of inorganic metal-halide layers, which are separated by long organic spacer molecules. They are described by their general chemical formula $L_2[ABX_3]_{n-1}BX_4$, where A is a small cation (e.g. methylammonium, formamidinium), B is a divalent metal cation (e.g. lead, tin), X is a halide anion (chloride, bromide, iodide), L is a long organic spacer molecule, and n is the number of octahedra that make up the thickness of the inorganic layer. The separation into few-atom thick inorganic layers yields strong quantum and dielectric confinement effects.[38] As a result, the exciton binding energies in 2D perovskites can be as high as several hundreds of meVs, which is around an order of magnitude larger than those found in bulk perovskites.[39–41] The excitonic character of the excited state is accompanied by an effective widening of the bandgap, an increase in the oscillator strength, and a narrowing of the emission spectrum.[40–42] The strongest confinement effects are observed for n = 1, where the excited state is confined to a single B-X-octahedral layer (see Figure 1a).

Light harvesting using 2D perovskites relies on the efficient transport of excitons and their subsequent separation into free charges.[43] This stands in contrast to bulk perovskites in which free charges are generated instantaneously thanks to the small exciton binding energy.[39] Particularly, with excitons being neutral quasi-particles, the charge extraction becomes significantly more challenging as they cannot be guided to the electrodes through an external electric field.[44] Excitons need to diffuse to an interface before the electron and hole can be efficiently separated into free charges.[45] On the other hand, for light emitting applications the spatial displacement is preferably inhibited, as a larger diffusion path increases the risk of encountering quenching sites which would reduce brightness. While charge transport in bulk perovskites has been studied in great detail, the mechanisms that dictate exciton transport in 2D perovskites remain elusive.[45] Moreover, it is unclear to what extent exciton transport can be controlled through variations in the perovskite composition.

Here, we report the direct visualization of exciton diffusion in 2D single-crystalline perovskites using time-resolved microscopy.[46] This technique allows us to follow the temporal evolution of a near-diffraction-limited exciton population with sub-nanosecond resolution and reveals the spatial and temporal exciton dynamics. We observe two different diffusion regimes. For early times, excitons follow normal diffusion, while for later times a subdiffusive regime emerges, which is attributed to the presence of trap states. Using the versatility of perovskite materials, we study the influence of the organic spacer on the diffusion dynamics of excitons in 2D perovskites. We find that between commonly used organic spacers (phenetylammonium, PEA, and butylammonium, BA), diffusivities and diffusion lengths can differ by one order of magnitude. We show that these changes are closely correlated with variations in the softness of the lattice, suggesting a dominant role for exciton-phonon coupling and exciton-polaron formation in the spatial dynamics of excitons in these materials. These insights provide a clear design strategy to further improve the performance of 2D perovskite solar cells and light emitting devices.

## Results

We prepare single crystals of n = 1 phenethylammonium lead iodine (PEA)$_2$PbI$_4$ 2D perovskite by drop-casting a saturated precursor solution onto a glass substrate,[47–49] as confirmed by XRD analysis and photoluminescence spectroscopy (see methods section for details). Using mechanical exfoliation, we isolate single-crystalline flakes of the perovskite and transfer these to microscopy slides. The single-crystalline flakes have typical lateral sizes of tens to hundreds of micrometers and are optically thick. The use of thick flakes provides a form of self-passivation that prevents the typical fast degradation of the perovskite in ambient conditions.

To measure the temporal and spatial exciton dynamics, we create a near-diffraction-limited exciton population using a pulsed laser diode ($\lambda_{ex}$ = 405 nm) and an oil immersion objective (N.A. = 1.3). The image of the fluorescence emission of the exciton population is projected outside the microscope with high magnification (330x), as illustrated in Figure 1b. By placing a scanning avalanche photodiode (20 μm in size) in the image plane, we resolve the time-dependent broadening of the population with high temporal and spatial resolution. Figure 1c shows the resulting map of the evolution in space and time of the fluorescence emission intensity of an exciton population in (PEA)$_2$PbI$_4$. The fluorescence emission intensity I(x,t) is normalized at each point in time to highlight the broadening of the emission spot over time. Each time-slice I(x,t$_c$) is well described by a Voigt function,[50] from which we can extract the variance $\sigma(t)^2$ of the exciton distribution at each point in time (Figure 1d). On a timescale of several nanoseconds, the exciton distribution broadens from an initial $\sigma(t=0)$ = 171 nm to $\sigma(t=10ns)$ = 448 nm, indicating fast exciton diffusion.

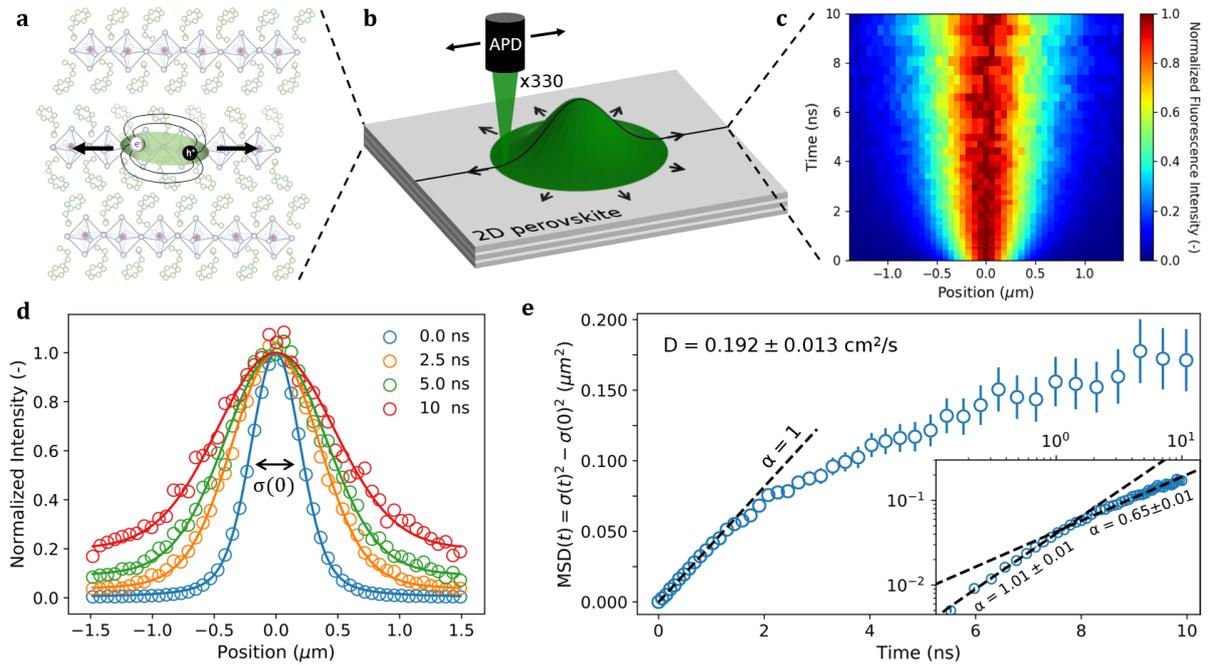

**Figure 1. (a)** Illustration of the (PEA)$_2$PbI$_4$ crystal structure, showing the perovskite octahedra sandwiched between the organic spacer molecules. **(b)** Schematic of the experimental setup. A near-diffraction limited exciton population is generated with a pulsed laser diode. The spatial and temporal evolution of the exciton population is recorded by scanning an avalanche photodiode through the magnified image of the fluorescence I(x,t). **(c)** Fluorescence emission intensity I(x,t) normalized at each point in time to highlight the spreading of the excitons. **(d)** Cross section of I(x,t) for different times t$_c$. **(e)** Mean-square-displacement of the exciton population over time. Two distinct regimes are present: First, normal diffusion with $\alpha$ = 1 is observed, which is followed by a subdiffusive regime with $\alpha$ < 1. The inset shows a log-log plot of the same data, highlighting the two distinct regimes. Reported errors represent the uncertainty in the fitting procedure for $\sigma(t)^2$.

To analyze the time-dependent broadening of the emission spot in more detail, we study the temporal evolution of the mean-square-displacement (MSD) of the exciton population, given by $MSD(t) = \sigma(t)^2 - \sigma(0)^2$. Taking the one-dimensional diffusion equation as a simple approximation, it follows that $MSD(t) = 2Dt^\alpha$, which allows us to extract the diffusivity $D$ and the diffusion exponent $\alpha$ from our measurement (see SI for derivation).[46,50] In Figure 1e we plot the MSD as a function of time. Two distinct regimes can be observed: For early times (t ≲ 1 ns) a fast linear broadening occurs with $\alpha = 1.01 \pm 0.01$, indicative of normal diffusion, while for later times (t ≳ 1 ns) the broadening becomes progressively slower with $\alpha = 0.65 \pm 0.01$, suggesting a regime of trap-state limited exciton transport (see SI for more details). The two regimes are clearly visible in the inset of Figure 1e, where different slopes correspond to different $\alpha$ values. From these measurements, a diffusivity of $0.192 \pm 0.013$ cm$^2$/s is found for (PEA)$_2$PbI$_4$.

The role of trap-states in perovskite materials is well studied and is generally attributed to the presence of imperfections at the surface of the inorganic layer.[51] These lower-energy sites lead to a subdiffusive behavior as a subpopulation of excitons becomes trapped. To test the influence of trap states, we have performed diffusion measurements in the presence of a continuous wave (CW) background excitation of varying intensity (Figure 2). The background excitation leads to a steady state population of excitons, which fill some of the traps and thereby reduce the effective trap density. To minimize the invasiveness of the measurement itself, the repetition rate and fluence were reduced to a minimum (see SI for details). In the absence of any background illumination, we find a strongly subdiffusive diffusion exponent of $\alpha = 0.48 \pm 0.02$. As the background intensity is increased, an increasing $\alpha$ is observed, indicative of trap state filling. Ultimately, a complete elimination of subdiffusion ($\alpha = 0.99 \pm 0.02$) is obtained at a background illumination power of 60 mW/cm². For comparison, this value corresponds roughly to a 2.5 Sun illumination. Additionally, we observe that the onset of the subdiffusive regime is delayed as more and more trap states are filled, as represented by the increasing t$_{split}$ parameter (see Figure 2b, bottom panel).

To gain theoretical insights and quantitative predictions concerning the observed subdiffusive behavior of excitons and its relation with trap state densities, we performed numerical simulations based on Brownian dynamics of individual excitons diffusing in a homogeneously distributed and random trap field (see SI for details). In addition, we developed a coarse-grained theoretical model based on continuum diffusion of the exciton concentration (see SI for details). The continuum theory predicts an exponential decay of the exciton diffusion coefficient,

$$\frac{1}{2}\frac{dMSD(t)}{dt} = D(t) = D\,exp\left(-\frac{D}{\lambda^2}t\right) \quad (1)$$

where $\lambda$ is the average distance between traps. The integral of this expression leads to

$$MSD(t) = 2\lambda^2\left[1 - exp\left(-\frac{D}{\lambda^2}t\right)\right], \quad (2)$$

which, as shown in Figure 2c, successfully reproduces both experimental and numerical results and allows us to accurately determine the value of the intrinsic trap state density, yielding $1/\lambda^2 = 22\ \mu m^{-2}$. The inset in Figure 2c shows the evolution of the effective trap state density $1/\lambda^2$ with increasing illumination intensity. We note that the exponential decay of equation (1) allows for a more intuitive characterization of *D(t)* by relating the subdiffusion directly to the trap density $1/\lambda^2$ rather than relying on the subdiffusive exponent $\alpha$ of a power law commonly used in literature.[46]

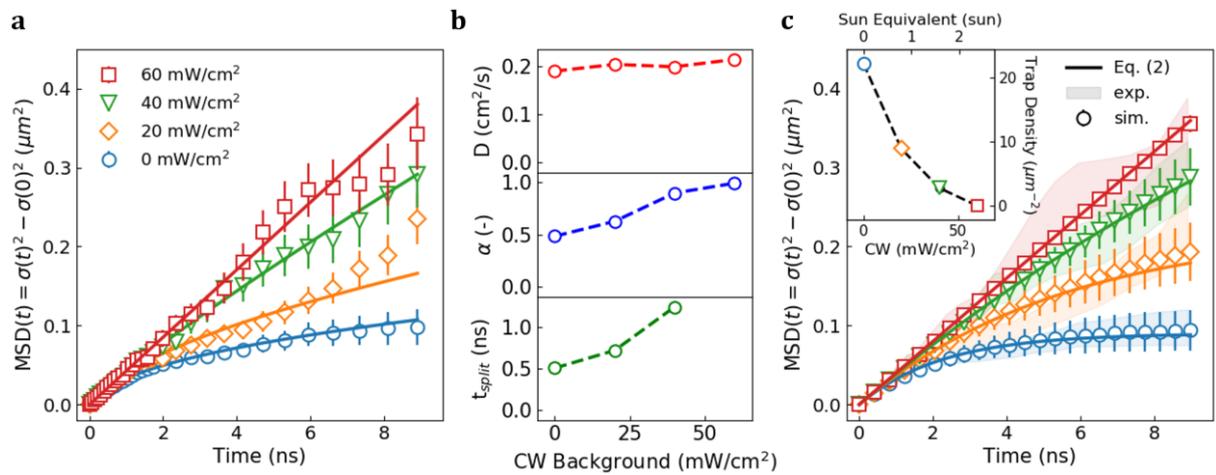

**Figure 2.** (a) Mean-square-displacement of the exciton population for different continuous wave (CW) background intensities. Experimental values are displayed with open markers, while the fit functions (equation S3), defined through the parameters *D*, $\alpha$, and $t_{split}$, are displayed as solid lines. (b) Diffusivity *D*, diffusion exponent $\alpha$, and the onset of subdiffusive regime $t_{split}$ extracted from fits in a). (c) Theoretical model (equation (2), solid lines), and numerical simulation (open markers) for exciton diffusion with different trap densities. Experimental values from a) are displayed as shaded areas for comparison. The inset shows the trap densities found with the simulations. Mirror axis of the inset is the sun equivalent of the background illumination intensity (AM1.5 Global and $E_{photon} > E_{bandgap}$).

Importantly, the early diffusion dynamics are unaffected by the trap density. This gives us direct access to the intrinsic exciton diffusivity of the material and allows us to compare the exciton diffusivity between perovskites of different composition. To explore compositional variations, we substitute phenetylamonium (PEA) with butylammonium (BA) - another commonly used spacer molecule for two-dimensional perovskites.[20,28,29,32,43,52–54]

Figure 3a displays the MSD of the (BA)$_2$PbI$_4$ perovskite, again showing the distinct transition from normal diffusion to a subdiffusive regime. However, as compared to (PEA)$_2$PbI$_4$, excitons in (BA)$_2$PbI$_4$ are remarkably less mobile, displaying a diffusivity of only $0.013 \pm 0.002$ cm$^2$/s, which is over an order of magnitude smaller than that of (PEA)$_2$PbI$_4$ with $0.192 \pm 0.013$ cm$^2$/s (green curve shown in Figure 3a for comparison). Taking the exciton lifetime into account, the difference in diffusivity results in a reduction in the diffusion length from $236 \pm 4\ nm$ for (PEA)$_2$PbI$_4$ to a mere $39 \pm 8\ nm$ for (BA)$_2$PbI$_4$ (see Figure 3b). These results indicate that the choice of ligand plays a crucial role in controlling the spatial dynamics of excitons in two-dimensional perovskites. We would like to note that the reported diffusion lengths follow the literature convention of diffusion lengths in one dimension, as it is the relevant length scale for device design. The actual two-dimensional diffusion length is greater by a factor of $\sqrt{2}$.

To understand the large difference in diffusivity between (PEA)$_2$PbI$_4$ and (BA)$_2$PbI$_4$, we take a closer look at the structural differences between these two materials. Changing the organic spacer can have a significant influence on the structural and optoelectronic properties of 2D perovskites. Specifically, increasing the cross-sectional area of the organic spacer distorts the inorganic lattice and reduces the orbital overlap between neighboring octahedra, which in turn increases the effective mass of the exciton.[55] Comparing the octahedral tilt angles of (PEA)$_2$PbI$_4$ and (BA)$_2$PbI$_4$, a larger distortion for the bulkier (PEA)$_2$PbI$_4$ (152.8°) as compared to (BA)$_2$PbI$_4$ (155.1°) is found.[56,57] The *larger* exciton effective mass in (PEA)$_2$PbI$_4$ would, however, suggest *slower* diffusion, meaning a simple effective mass picture for free excitons cannot explain the observed trend in the diffusivity between (PEA)$_2$PbI$_4$ and (BA)$_2$PbI$_4$.

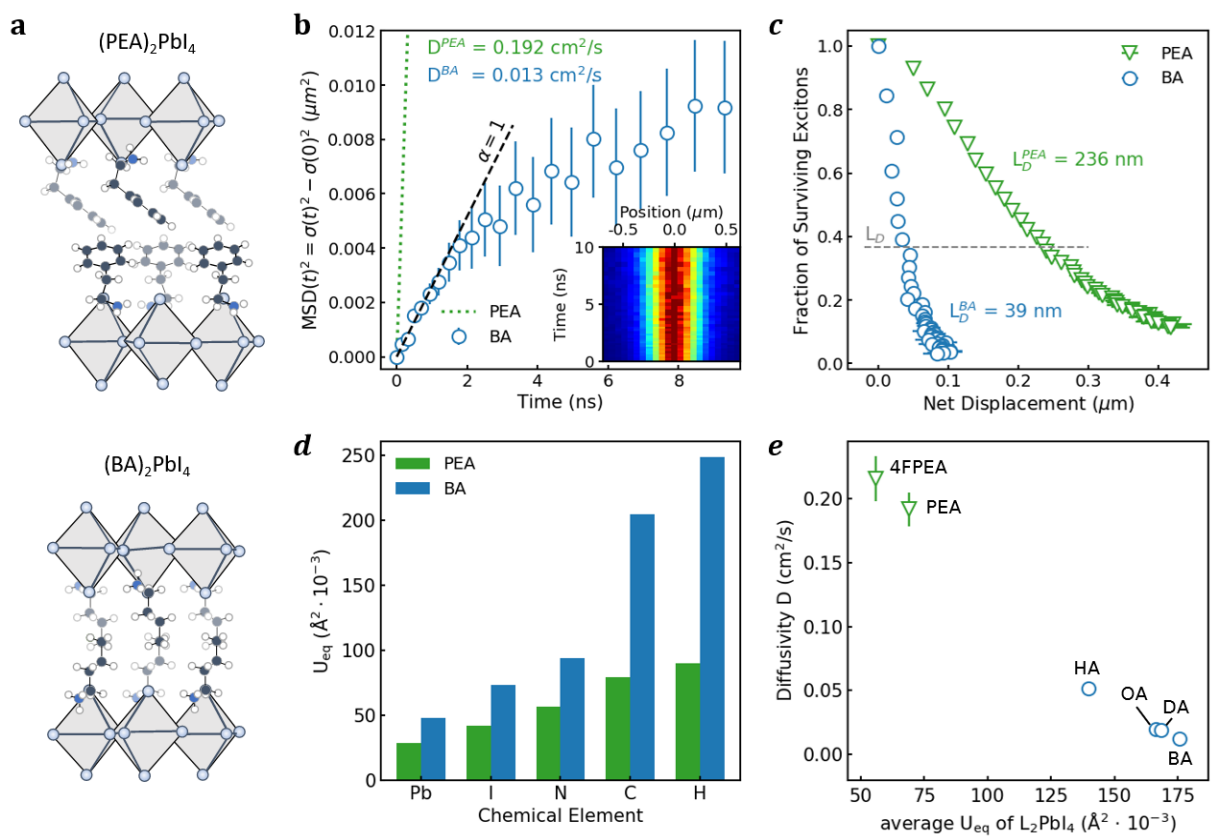

Figure 3. (a) $(PEA)_2PbI_4$ and $(BA)_2PbI_4$ crystal structure.[56,57] (b) Mean-square-displacement of exciton population over time for $(PEA)_2PbI_4$ (dotted line) and $(BA)_2PbI_4$ (circles). Inset shows the normalized fluorescence emission intensity I(x,t) for $(BA)_2PbI_4$. (c) Fractions of surviving excitons (extracted from lifetime data in Figure S7) vs. net spatial displacement $\sqrt{MSD(t)}$ of excitons for $(PEA)_2PbI_4$ (triangles) and $(BA)_2PbI_4$ (circles). (d) Average atomic displacement $U_{eq}$ of the chemical elements in $(PEA)_2PbI_4$ and $(BA)_2PbI_4$. Data was extracted from previously published single crystal x-ray diffraction data.[56,57] (e) Diffusivity D vs. average atomic displacement $U_{eq}$ for different organic spacers: 4-fluoro-phenethylammonium (4FPEA),[58] phenethylammonium (PEA),[56] hexylammonium (HA),[57] octylammonium (OA),[59] decylammonium (DA),[59] Butylammonium (HA).[57]

Rather than dealing with free excitons, a number of studies have pointed at the importance of strong exciton-phonon coupling and the formation of exciton-polarons in perovskite materials.[35,37,60,61] In the presence of an exciton, the soft inorganic lattice of the perovskite can be easily distorted through coupling with phonons, leading to the formation of polarons. As compared to a free exciton, an exciton-polaron exhibits a larger effective mass and, consequently, a lower diffusivity. The softer the lattice, the larger the distortion, and the heavier the polaron effective mass will be.[62] The correct theoretical description of the polaron in 2D perovskites is the subject of ongoing debate, though the current consensus is that the polar anharmonic lattice requires a description beyond conventional Frohlich theory.[60,61,63]

To investigate the potential role of polaron formation on exciton diffusion, we first quantify the softness of the lattices of both $(PEA)_2PbI_4$ and $(BA)_2PbI_4$ by extracting the atomic displacement parameters from their respective single crystal x-ray data.[64] The atomic displacement of the different atoms of both systems are summarized in Figure 3d, showing distinctly larger displacements for $(BA)_2PbI_4$ as compared to $(PEA)_2PbI_4$ in both the organic and inorganic sublattice.[56,57] The increased lattice rigidity for $(PEA)_2PbI_4$ can be attributed to the formation of an extensive network of pi-hydrogen bonds and a more space-filling nature of the aromatic ring, both of which are absent in the aliphatic BA spacer molecule. Qualitatively, a stiffening of the lattice reduces the exciton-phonon coupling and would explain the observed higher diffusivity in $(PEA)_2PbI_4$ as compared to $(BA)_2PbI_4$. In addition to a softer lattice, we find that $(BA)_2PbI_4$ exhibits a larger exciton-phonon coupling strength as compared to $(PEA)_2PbI_4$, as confirmed by analyzing the temperature-dependent broadening of the photoluminescence linewidth of the two materials (see also Figure S8-S10 and Table S2).[65] The combined effect of a softer lattice and a larger exciton-phonon coupling strength in $(BA)_2PbI_4$ as compared to $(PEA)_2PbI_4$ promotes the formation of exciton-polarons with heavier effective masses, consistent with the observed trend in the diffusivity.

To further test the correlation between lattice softness and diffusivity, we have performed measurements on a wider range of two-dimensional perovskites with different organic spacers. In

Figure 3e, we present the diffusivity as a function of average atomic displacement for each of the different perovskite unit cells. Across the entire range of organic spacers, a clear correlation between the diffusivity and the lattice softness is found, consistent with exciton-polaron formation as the dominant parameter in determining the spatial dynamics of the excited state.

Finally, we look at the temperature dependence of the diffusivity, which can give us critical insights into the nature of the polaron. When long-range deformations of the lattice are dominant, the exciton-polaron extends across multiple lattice sites and is categorized as a large polaron. The diffusion of large polarons occurs coherently and decreases with increasing temperature ($\partial D/\partial T < 0$), resembling that of band-like free exciton motion, although with a strongly increased effective mass. Contrarily, in the case of dominant short-range lattice deformations, small polarons are formed that localize within the unit cell of the material. The motion of small polarons occurs through incoherent site-to-site hopping and increases with temperature ($\partial D/\partial T > 0$). In both $(PEA)_2PbI_4$ and $(BA)_2PbI_4$, a clear negative scaling of the diffusivity with temperature is observed ($\partial D/\partial T < 0$, see Figure 4), consistent with the formation of large exciton-polarons and coherent exciton transport through the 2D plane.

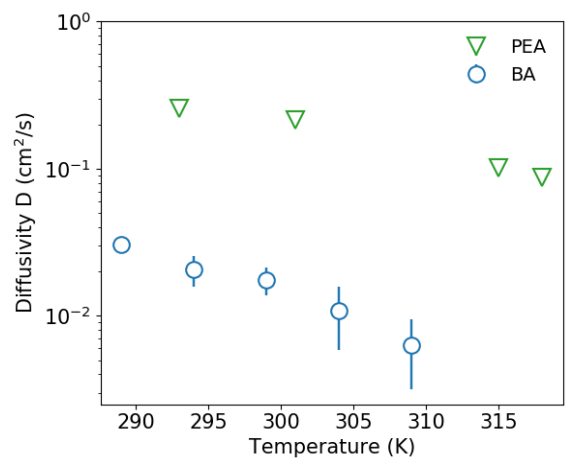

**Figure 4. Temperature dependent diffusivity in $(PEA)_2PbI_4$ (triangles) and $(BA)_2PbI_4$ (circles).**

## Discussion

Given the apparent importance of exciton-phonon coupling in the spatial dynamics of excitons in 2D perovskites, structural rigidity becomes a critical design parameter in these systems. Taking into account the close correlation between diffusivity and the atomic displacement, this parameter space can be readily explored using available x-ray crystal structure data for many 2D perovskite analogues. While the influence of the organic spacer is expected to be particularly strong in the class of n = 1 2D perovskites, we have observed consistent trends in the n = 2 analogues. Indeed, just like in n = 1, in n = 2 the use of the PEA cation yields higher diffusivities than for BA (see supporting information). Similarly, the interstitial formamidinium (FA) cation in n = 2 yields higher diffusivity than the methylammonium (MA) cation, consistent with the trend in the atomic displacement parameters.

From a technological perspective, structural rigidity may play a particularly important role in light emitting devices. Long exciton diffusion lengths in light emitting applications can act detrimentally on device performance, as it increases the possibility of encountering a trapping site. From an exciton-polaron perspective, this suggests soft lattices are preferred. At the same time though, Gong et al. highlighted the role of structural rigidity in improving the luminescence quantum yield through a reduced coupling to non-radiative decay pathways.[64,66] A trade-off therefore exists in choosing the optimal rigidity for bright emission. Meanwhile, for light harvesting applications, long diffusion lengths are essential for the successful extraction of excitons. While strongly excitonic 2D perovskites are generally to be avoided due to the penalty imposed by the exciton binding energy, improving the understanding of the spatial dynamics of the excitonic state may help mitigate this negative impact of the thinnest members of the 2D perovskites in solar harvesting.

## Acknowledgements

This work has been supported by the Spanish Ministry of Economy and Competitiveness through The "María de Maeztu" Program for Units of Excellence in R&D (MDM-2014-0377). M.S. acknowledges the financial support of a fellowship from "la Caixa" Foundation (ID 100010434). The fellowship code is LCF/BQ/IN17/11620040. M.S. has received funding from the European Union's Horizon 2020 research and innovation program under the Marie Skłodowska-Curie grant agreement No. 713673. F.P. acknowledges support from the Spanish Ministry for Science, Innovation, and Universities through the state program (PGC2018-097236-A-I00) and through the Ramón y Cajal program (RYC-2017-23253), as well as the Comunidad de Madrid Talent Program for Experienced Researchers (2016-T1/IND-1209). N.A., M.M. and R.D.B. acknowledges support from the Spanish Ministry of Economy, Industry and Competitiveness through Grant FIS2017-86007-C3-1-P (AEI/FEDER, EU). E.P. acknowledges support from the Spanish Ministry of Economy, Industry and Competitiveness through Grant FIS2016-80434-P (AEI/FEDER, EU), the Ramón y Cajal program (RYC-2011- 09345) and the Comunidad de Madrid through Grant S2018/NMT-4511 (NMAT2D-CM). S.P. acknowledges financial support by the VILLUM FONDEN via the Centre of Excellence for Dirac Materials (Grant No. 11744).

## Author contributions

M.S. and F.P. designed this study. M.S. led the experimental work and processing of experimental data. M.S. set up the diffusion measurement technique with the assistance of T.J.L., and S.W.W. A.J.M. and M.S. performed temperature dependent measurements. M.S. and A.J.M. prepared perovskite materials. N.A., M.M., and R.D-B. performed theoretical and numerical modelling of exciton transport. M.S, F.P., S.P., and E.P. provided the theoretical interpretation of the intrinsic exciton transport. F.P. supervised the project. M.S. and F.P. wrote the original draft of the manuscript. All authors contributed to reviewing the manuscript.

Supporting information for:

# Exciton diffusion in two-dimensional metal-halide perovskites

Michael Seitz[1,2], Alvaro J. Magdaleno[1,2], Nerea Alcázar-Cano[1,3], Marc Meléndez[1,3], Tim J. Lubbers[1,2], Sanne W. Walraven[1,2], Sahar Pakdel[4], Elsa Prada[1,2], Rafael Delgado-Buscalioni[1,3], and Ferry Prins[1,2]

*1. Condensed Matter Physics Center (IFIMAC), Autonomous University of Madrid, 28049 Madrid, Spain*

*2. Department of Condensed Matter Physics, Autonomous University of Madrid, 28049 Madrid, Spain*

*3. Department of Theoretical Condensed Matter Physics, Autonomous University of Madrid, 28049 Madrid, Spain*

*4. Department of Physics and Astronomy, Aarhus University, 8000 Aarhus C, Denmark*

1. Sample preparation
2. Exciton diffusion measurements
3. Trap state dynamics
4. Determination of diffusion length
5. Temperature dependent photoluminescence linewidth
6. Diffusivities for different chemical compositions
7. References

## 1. Sample Preparation

**Chemicals:** Chemicals were purchased from commercial suppliers and used as received:

*$MX_2$:* lead(II) iodide ($PbI_2$) (Sigma Aldrich, 900168-5G).

*LX:* phenethylammonium iodide (PEAI) (Sigma Aldrich, 805904-25G), n-butylammonium iodide (BAI) (Sigma Aldrich, 805874-5G), n-octylammonium iodide (OAI) (Greatcell Solar Materials, MS105500-5), 4-Fluoro-Phenethylammonium iodide (4FPEAI) (Greatcell Solar Materials, MS100720-05)

*AX:* formamidinium iodide (Greatcell Solar Materials, MS150000-05), methylammonium iodide (Sigma Aldrich, 793493-5G).

*Solvents:* γ-butyrolactone (Sigma Aldrich, B103608-500G)

n-hexylammonium iodide (HAI) and n-decylammonium iodide (DAI) were not purchased directly from a commercial supplier but were synthesized by reacting the amine species with hydriodic acid (HI).

*L - amines*: hexylamine (Sigma Aldrich, 219703-100ML), decylamine (Sigma Aldrich, D2404-5G)

*Acid*: hydriodic acid (HI) (Sigma Aldrich, 752851-25G)

**Abbreviations:** For the ease of writing several abbreviations are used, which are summarized here: phenethylammonium (PEA), butylammonium (BA), hexylammonium (HA), octylammonium (OA), decylammonium (DA), 4-Fluoro-Phenethylammonium (4FPEA), lead (Pb), iodide (I), methylammonium (MA), formamidinium (FA), two-dimensional (2D)

**Synthesis:** Layered perovskites, with the exception of $(HA)_2PbI_4$ and $(DA)_2PbI_4$, were synthesized under ambient laboratory conditions following the over-saturation techniques reported previously.[1–3] In a nutshell, the precursor salts LI, $PbI_2$, and AI were mixed in a stoichiometric ratio (2:1:0 for n = 1 and 2:2:1 for n = 2) and dissolved in γ-butyrolactone. The solution was heated to 70 °C and more γ-butyrolactone was added (while stirring) until all the precursors were completely dissolved. The resulting solutions were heated to 70°C and the solvent was left to evaporate. After 2-3 days, millimeter sized crystals formed in the solution, which was subsequently cooled down to room temperature. For this study, we drop cast some of the remaining supersaturated solution on a glass slide, heated it up to 50 °C with a hotplate and after the solvent was evaporated, crystals with crystal sizes of up to several hundred microns were formed. The saturated solution can be stored and re-used to produce freshly grown 2D perovskites within several minutes. We would like to note that drop cast n = 2 solutions form several crystals with different n values. However, n = 2 crystals can be easily isolated during the exfoliation (see next section) and the formation of n = 2 can be favored by preheating the substrate to 50°C before drop casting.

(HA)$_2$PbI$_4$ and (DA)$_2$PbI$_4$ were synthesized by dissolving PbI$_2$ (100 mg) in HI (800 μl) through heavy stirring and heating the solution to 90°C. After PbI$_2$ was completely dissolved a stoichiometric amount of the amine was added dropwise to the solution.

**Exfoliation:** The perovskite crystals of the thin film were mechanically exfoliated using the Scotch tape method (Nitto SPV 224). The exfoliation guarantees a freshly cleaved and atomically flat surface area for inspection, which is crucial to avoid emission from edge states and guarantee direct contact with the glass substrate. After several exfoliation steps, the crystals were transferred on a glass slide and were subsequently studied through the glass slide with a 100x oil immersion objective (Nikon CFI Plan Fluor, NA = 1.3). A big advantage of this technique is that the perovskites are encapsulated through the glass slide from one side and by the bulk of the crystal from the other side. It is important to use thick crystals to guarantee good "self-encapsulation" and prevent premature degradation of the perovskite flakes to affect the measurement.[3]

**X-ray diffraction (XRD):** XRD was performed with a PANanaltical X'Pert PRO operating at 45 kV and 40 mA using a copper radiation source (λ = 1.5406 Å). The polycrystalline perovskite films were prepared by drop casting the saturated perovskite solutions on a silicon zero diffraction plate.

We use XRD measurements to extract the lattice spacing d of the inorganic layers of 2D perovskites by extracting the diffraction peak positions $\theta_n$ and using Bragg's law $n\lambda = 2d \cdot \sin(\theta_n)$. Figure S1,2 show the XRD pattern of polycrystalline (PEA)$_2$PbI$_4$ and (BA)$_2$PbI$_4$ films, respectively. The extracted lattice spacings are consistent with previously reported values of these materials.[4,5] Table S1 summarizes the XRD data of the perovskite crystals used in this study.

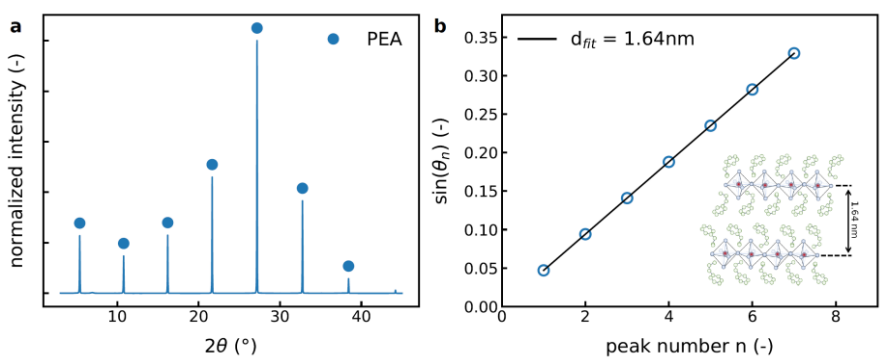

**Figure S1.** (a) XRD pattern of (PEA)$_2$PbI$_4$ (b) In accordance with Bragg's law sin($\theta_n$), where $\theta_n$ are the diffraction peak angles, follows a linear behavior and reveals a spacing of 1.64 nm between inorganic layers.

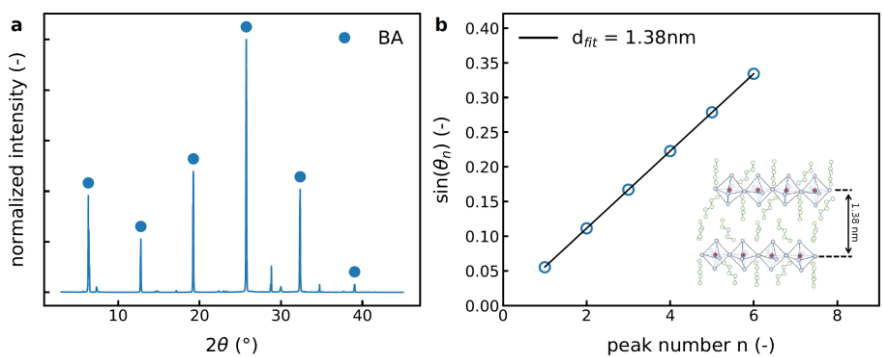

**Figure S2.** (a) XRD pattern of (BA)$_2$PbI$_4$ (b) In accordance with Bragg's law sin($\theta_n$), where $\theta_n$ are the diffraction peak angles, follows a linear behavior and reveals a spacing of 1.38 nm between the inorganic layers.

**Table S1.** Summary of XRD measurements for perovskites in this paper. Table shows the lattice spacing d obtained in this study, reported lattice spacing $d_{lit}$ from literature, and chemical drawing of the used organic spacer L and cation A of the perovskite (L$_2$[APbI$_3$]$_{n-1}$PbI$_4$).

|  | $d$ (nm) | $d_{lit}$ (nm) | organic spacer L | cation A |
|---|---|---|---|---|
| (PEA)$_2$PbI$_4$ | 1.64 | 1.65[4] | | - |
| (4FPEA)$_2$PbI$_4$ | 1.66 | 1.67[6] | | - |
| (BA)$_2$PbI$_4$ | 1.38 | 1.38[5] | | - |
| (HA)$_2$PbI$_4$ | 1.64 | 1.64[5] | | - |
| (OA)$_2$PbI$_4$ | 1.88 | 1.87[7] | | - |
| (DA)$_2$PbI$_4$ | 2.13 | 2.13[7] | | - |
| (BA)$_2$(MA)Pb$_2$I$_7$ | 1.97 | 1.96[8] | | |
| (BA)$_2$(FA)Pb$_2$I$_7$ | 1.97 | 1.96[8] | | |

## 2. Exciton Diffusion Measurements

**Experimental setup:** Exciton diffusion measurements were measured following the same procedure as reported previously.[9,10] In short, a near diffraction limited exciton population was created using a 405 nm laser (PicoQuant LDH-D-C-405, PDL 800-D) and a 100x oil immersion objective (Nikon CFI Plan Fluor, NA = 1.3). Fluorescence of the exciton population was then imaged with a total 330x magnification onto an avalanche photodiode (APD, Micro Photon Devices PDM) with a detector size of 20 µm. The laser and APD were synchronized using a timing board for time correlated single photon counting (Pico-Harp 300). The APD was capturing an effective area of around 60 x 60 nm (= 20 µm / 330). The APD was scanned through the middle of the exciton population in 60 or 120 nm steps, recording a time trace in every point. To minimize the degradation of the perovskites through laser irradiation, the perovskite flakes were scanned using an x-y-piezo stage (MCL Nano-BIOS 100), covering an area of 5 x 5 µm. Diffusion measurements were performed with a 40 MHz laser repetition rate and a laser fluence of 50 nJ/cm² unless stated otherwise. The time binning of the measurement was set to 4 ps before software binning was applied.

For the temperature dependent measurements, the temperature was varied with a silicon heater mat (RS PRO, 245-499), using a PID temperature controller (Dwyer Instruments, Series 16C-3) connected to a type K thermocouple (Labfacility, Z2-K-1M) for feedback control. Here, a silicon heater mat was chosen over the Peltier element as a Peltier element expands during the heating process and causes mechanical vibrations that lead to drift.

**Derivation of the mean-square-displacement (MSD) from the diffusion equation:** For the theoretical model we followed the derivation by Akselrod et al.[10] For convenience we provide a summary here. The temporal and spatial exciton dynamics can be described by the general diffusion equation:

$$\dot{n}(x,t) = g(x,t) - \frac{n(x,t)}{\tau_r} - \frac{n(x,t)}{\tau_{nr}} + D\nabla^2 n(x,t),$$

where, $n(x,t)$ is the exciton density distribution, $g(x,t)$ the exciton generation term, $\tau_r$ the lifetime of radiative decay and $\tau_{nr}$ of non-radiative decay, $D$ the diffusion constant and $\nabla$ the gradient. Please note that higher order processes, such as exciton-exciton annihilation, are not taken into account because only low laser fluences are used in this study. The generation term can be assumed to be 0 for times larger than 0: $g(t > 0) = 0$ and can be accounted for in the initial conditions as $n(x, 0) = g(x)$. In addition, the diffusion equation can be separated into a single dimension, because diffusion in x- and y-direction are uncorrelated, and our measurement takes a slice along a single Cartesian direction. For the sake of simplicity the radiative and non-radiative decay constants can be summarized in a single term as $\tau =$ $\left(\frac{1}{\tau_r} + \frac{1}{\tau_{nr}}\right)^{-1} = \frac{\tau_r \tau_{nr}}{\tau_r + \tau_{nr}}$, without any loss of generality. The simplified one-dimensional diffusion equation then becomes

$$\dot{n}(x,t) = -\frac{n(x,t)}{\tau} + D\frac{d^2 n(x,t)}{dx^2},$$

where $n(x,t)$ is the exciton density distribution, $\tau$ is the lifetime of the excitons and $D$ is the diffusivity. The general solution with constant diffusivity D, constant decay rate $\tau$, and an initial exciton distribution $n(x,0)$ is:

$$n(x,t) = e^{-\frac{t}{\tau}} \frac{1}{\sqrt{4\pi Dt}} \int_{-\infty}^{\infty} n(x_0, 0) e^{-\frac{(x-x_0)^2}{4Dt}} dx_0 = e^{-\frac{t}{\tau}}[n(x,0) * G(x,t)], \quad \text{(S1)}$$

where $G(x,t) = \frac{1}{\sqrt{4\pi Dt}} e^{-\frac{x^2}{4Dt}}$. This means that the exciton density distribution is a convolution of the initial exciton distribution $n(x,0)$ and a Gaussian function $G(x,t)$, with a variance of $\sigma(t)^2 = 2Dt$, multiplied with the temporal decay $e^{-\frac{t}{\tau}}$ of the population. From equation (S1) one can see that the lifetime $\tau$ of the excitons does not influence the broadening of the exciton population, allowing us to continue with the normalized quantity $\hat{n}(x,t) = \frac{n(x,t)}{e^{-\frac{t}{\tau}}}$. To approximate the initial exciton density distribution $n(x,0)$ a Lorentzian function is used, as it best describes the observed initial exciton distribution:[10]

$$\hat{n}(x,t) = n(x,0) * G(x,t) = L(x) * G(x,t)$$

In the experiment, the measured emission pattern $I(x,t)$ is a spatial convolution of the exciton distribution $\hat{n}(x,t)$, the APD detector $f_d(x)$ and the point spread function (PSF) $f_{psf}(x)$ of the optical system:

$$I(x,t) = \hat{n}(x,t) * f_d(x) * f_{psf}(x) = L(x) * G(x,t) * f_d(x) * f_{PSF}(x) \quad \text{(S2)}$$

The APD detector size of 20 µm, which when imaged onto the sample with a 330x magnification becomes 60 nm, is small when compared to the laser spot size. Therefore, the influence of the APD detector size is negligible and equation (S2) can be rewritten as

$$I(x,t) = L(x) * G(x,t) * f_{PSF}(x)$$

Approximating the PSF of the microscope with a Gaussian function and knowing that a convolution of two Gaussians is a gaussian as well, allows us to further simplify the equation:

$$I(x,t) = L(x) * G(x,t) * f_{PSF}(x) = L(x) * \tilde{G}(x,t) \propto L(x) * e^{-\frac{x^2}{2\sigma_I^2(t)}}$$

Since the variance of convoluted Gaussians is additive we can write $\sigma_I^2(t) = 2Dt + \sigma_{PSF}^2$, where $\sigma_I^2(t)$ is the variance of $\tilde{G}(x,t)$, 2Dt is the variance due to the diffusion term $G(x,t)$, and $\sigma_{PSF}^2$ is the variance of the point spread function of the microscope, which is constant over time.

Fitting the experimental data $I(x,t)$ with a Voigt function (convolution of Lorentzian and Gaussian function), allows the extraction of the full-width-half-maximum (fwhm) of the Lorentzian part $L(x)$ and the variance $\sigma_I^2(t)$ of the Gaussian part $\tilde{G}(x,t)$. As a result, the mean-square-displacement (MSD) of an exciton can be calculated by subtracting the variance of the Gaussian ($\sigma_I^2(t)$) by the variance at time 0 ($\sigma_I^2(0)$):

$$MSD(t) = \sigma_I^2(t) - \sigma_I^2(0) = 2Dt + \sigma_{PSF}^2 - \sigma_{PSF}^2 = 2Dt$$

Consequently, the diffusivity D can be extracted by measuring the dime-dependent variance $\sigma_I^2(t)$ of the exciton distribution.

To extract the MSD of the exciton population for different times t, we fit the population with a Voigt distribution. A Voigt function is completely defined through five parameters: The fwhm of the Lorentzian part $\Gamma$, the variance of the Gaussian part $\sigma^2$, the center of the distribution $\mu$, the proportionality factor A, and the offset c due to, for example, background noise: $V(x) = A\left(\frac{\Gamma/2}{(x-\mu)^2+\Gamma^2/4} * e^{-\frac{(x-\mu)^2}{2\sigma^2}}\right) + c$. $\Gamma$, $\mu$ and c are constant for all times, while $\sigma$ and A change over time. In our fitting procedure we fit the data such that for every time-slice we allowed different $\sigma_k$ and $A_k$ parameters (indexed here with k), while $\Gamma$, $\mu$ and c were shared and constant for all times. As a result, the fitting includes 3 + 2k parameters. The fitting is performed in python with the optimize.curve_fit function of the scipy package. Although we are only interested in the width of the exciton population ($\sigma_k$), we did fit the non-normalized data as this guarantees that every data point of I(x,t) is weighted according to its signal-to-noise (S/N) ratio. If the normalized data would be used, data points for late times would get weighted the same as early time points in the fitting procedure, despite a much inferior S/N ratio of data points collected at later times.

The S/N of fluorescence lifetime data significantly decreases for later times due to the exponential behavior. As a result, we choose to apply a non-linear binning to our data before the fitting procedure. Meaning at early times fewer bins are added together ($I_{binned} = \sum_{k=i}^{i+n_{early}} \frac{I_k}{n_{early}}$) than for later times ($I_{binned} = \sum_{k=i}^{i+n_{late}} \frac{I_k}{n_{late}}$). Hence, $n_{early} < n_{late}$. The number of bins is chosen according to the following function: $n_k = int(0.3k^{1.5} + 16) \rightarrow [16, 16, 16, 17, 18, 19, 20, 21, 22, 24 ...]$. The time bin size before binning was 4 ps.

Binning improves the S/N ration of the data. With non-linear binning, the $(S/N)_k$ for every time-slice improves differently. As a result, every time-slice is weighted with the number of bins $n_k$ to correctly account for the improved $(S/N)_k$ due to binning. This multiplication with $n_k$ has an influence on the actual offset parameter c, that needs to be fitted to each time-slice

k. After non-linear binning, every time-slice needs to be fitted with a local offset of $c_k = c \cdot n_k$, with c being the real/global offset of the data. Following this procedure, we are able to extract the variance $\sigma_k^2$ ($\sigma_I^2(t)$ in the previous section) for every time-slice, which are then used to generate the MSD ($= \sigma_I^2(t) - \sigma_I^2(0) = \sigma_k^2 - \sigma_0^2$) vs. time plots, that allowed the extraction of the diffusion parameters D, $\alpha$ and $t_{split}$.

We would like to note that the shortest lifetime of some perovskites is close to the width of the pulse-width of the pulsed laser diode. To minimize the influence of the laser and justify the assumption of negligible exciton generation during our measurements ($g(t > 0) = 0$) we only analyze the diffusion data 250 ps after the maximum of the photoluminescence lifetime data and define it as the new t = 0.

In the previous paragraphs, we derived that the MSD of normal diffusion in one-dimension to be: $MSD(t) = 2Dt$. However, diffusion in disordered media, where the diffusivity is not constant for all times, is better described by introducing the diffusion exponent $\alpha$: $MSD(t) = 2Dt^\alpha$.[9–11] With the diffusion exponent $\alpha$ one can describe normal diffusion ($\alpha = 1$), superdiffusion ($\alpha > 1$, e.g. ballistic transport) and subdiffusion ($\alpha < 1$, e.g. through trapping of excitons).

As described in the main text the MSD(t) of excitons in 2D perovskites shows two distinct diffusion regimes: First, a linear behavior of normal diffusion ($2Dt^\alpha$, $\alpha = 1$), which is followed by a second subdiffusive regime ($2Dt^\alpha$, $\alpha < 1$). We fit the two regimes simultaneously with the following fit function:

$$MSD(t) = \begin{cases} 2Dt + c & for \ t \leq t_{split} \\ 2D\left[\left(t - t_{split} + t_0\right)^\alpha + t_{split} - t_0^\alpha\right] + c & for \ t > t_{split} \end{cases} \quad (S3)$$

with the fit parameters D, $\alpha$, c, and $t_{split}$. c is generally small and is introduced to avoid overweighing the first datapoint at time t = 0. $t_0 = \alpha^{\frac{1}{1-\alpha}}$ was introduced to make MSD(t) continuous in value (excitons move continuously) and slope (speed of excitons changes continuously) at time $t_{split}$. By fitting our experimental data with equation (S3) we were able to extract the diffusivity D, diffusion exponent $\alpha$, and the onset of subdiffusion $t_{split}$ from our measurements. We would like to note that fitting only the first linear regime of normal diffusion with $2Dt^\alpha$ yields almost identical D values and $\alpha$ values of around 1.

**Brownian modelling of trap states**

We have performed Brownian dynamics simulations of a single exciton diffusing in a field of traps, representing ideal (non-interacting) excitons in the dilute limit carried out in experiments.

In these simulations, an exciton diffuses freely until it finds a trap, where it just stops. Free diffusion is modelled using the standard stochastic differential equation for Brownian motion in the Itô interpretation. If *r(t)* is the position of the exciton in the plane at time *t*, its displacement *Δr* over a time *Δt* is given by,

$$\Delta \boldsymbol{r} = \sqrt{2D}\, d\boldsymbol{W},$$

where *D* is the free-diffusion coefficient and *dW* is taken from a Wiener process, such that $\langle d\boldsymbol{W} d\boldsymbol{W}\rangle = \Delta t$. Traps were scattered throughout the plane following a uniform random distribution. The exciton is considered to be trapped as soon its location gets closer than $R_{trap}$ = 1.2 nm to the trap center. The value was taken from estimations of the exciton Bohr radius and corresponds to a trap area of 1.44 nm$^2$.[12] In any case, in the dilute regime, the diffusion is not sensitive to the trap size $R_{trap}$, because the trap radius is much smaller than the average separation between traps, $R_{trap} \ll \lambda$. To numerically integrate the equation of motion, we used a simple second-order-accurate modification of the well-known Euler Maruyama algorithm: the BAOAB-Limit method.[13] Trajectories were computed for many independent excitons and the data was averaged to determine the MSD as a function of time.

**Derivation of continuum model:**

Besides studying the exciton dynamics using Brownian dynamics simulations, we derive a continuum model for the exciton diffusion in a plane having a uniform random distribution of traps. This coarser model solves the equation for the field of exciton concentration, or equivalently, for the probability field *c(r,t)* of finding an exciton at location *r* at time *t*. The resulting partial differential equation for *c(r,t)* is numerically solved using a finite difference scheme in a rectangular mesh of size *h*. Thus, in this description, *r* corresponds to the center of a given control cell of the mesh, whose area *h²* is much larger than the trap area $s_0 = \pi R_{trap}^2$, but yet much smaller than the system size *L*. In other words, $R_{trap} \ll h \ll L$. To model the spatial distribution (mean-square-displacement), we consider two kinds of excitons: mobile excitons which freely diffuse with a diffusion constant *D* and trapped excitons which stand still. Note that the temporal dynamics, such as the radiative lifetime of excitons, can be neglected since we are only interested in the spatial dynamics. The total concentration of excitons is just

$$C_{tot}(\boldsymbol{r},t) = c(\boldsymbol{r},t) + c_t(\boldsymbol{r},t),$$

where *c(r,t)* is the concentration of mobile excitons and *c$_t$(r,t)* that of the trapped ones. Any increase in trapped excitons is due to a loss in mobile excitons, $\delta c_t(\boldsymbol{r},t) = -\delta c(\boldsymbol{r},t)$. The concentration of mobile excitons will obey a reaction-diffusion equation, with a free diffusion term ($D\nabla^2 c$) and sink or loss rate term, *s(r,t)*,

$$\partial_t c(\mathbf{r}, t) = D\nabla^2 c(\mathbf{r}, t) - s(\mathbf{r}, t).$$

The equation for trapped excitons is then simply, $\partial_t c_t(\mathbf{r}, t) = s(\mathbf{r}, t)$, and that for the total number of excitons,

$$\partial_t C_{tot}(r, t) = \nabla \cdot (D\nabla c(\mathbf{r}, t)) \tag{S4}$$

which is explicitly written in conservative form: note that the total exciton flux is just the diffusive flux of mobile excitons $\mathbf{j} = -D\nabla c$.

As we show below, the exciton's mean square displacement is directly given by equation (S4). But, at this stage one needs to model the loss rate *s(r,t)*, provided that the surface fraction of trap *p* is fixed, and the trap area is $s_0$. Recall that the average distance between traps is λ so, $p = s_0/\lambda^2$ and the trap density is $p/s_0$. The loss rate has two contributions. First, over a time lapse *Δt*, the excitons located at the cell *r* (whose number is *c(r,t) h²*) will diffuse an area $A = \gamma D_0 \Delta t$ (where γ represents a dimensionless proportionality factor), and a fraction *p A/$s_0$* of them will fall in a trap. The second contribution to *s(r,t)* is due to the flux of excitons across the boundaries of the cell *r*. The number of excitons crossing the boundaries of the cell *r* (to or from another cell) is just

$$\int \mathbf{j} \cdot n d\mathbf{r} = \int \nabla \cdot (D\nabla c) d\mathbf{r}^2 \approx h^2 D \nabla^2 c.$$

Here, we have used the Gauss divergence theorem and the fact that the cell size *h* is infinitesimal. Again, a fraction *p* of these excitons will also fall in a trap. In summary, the sink (loss rate) is given by

$$s(\mathbf{r}, t) = \gamma p D \nabla^2 c(\mathbf{r}, t) + \gamma \frac{p}{s_0} D c(\mathbf{r}, t),$$

and the resulting equation for the mobile excitons is

$$\partial_t c(\mathbf{r}, t) = D(1 - \gamma p)\nabla^2 c(\mathbf{r}, t) - \gamma p \frac{D}{s_0} c(\mathbf{r}, t). \tag{S5}$$

This equation has a particularly simple analytical solution in Fourier space,

$$\partial_t \tilde{c}(\mathbf{q}, t) = -D(1 - \gamma p)\mathbf{q}^2 \tilde{c}(\mathbf{q}, t) - \gamma \frac{D_0}{\lambda^2} \tilde{c}(\mathbf{q}, t),$$

with $\tilde{c}(\mathbf{q}, t) = \tilde{c}(t) e^{-i\mathbf{q} \cdot \mathbf{r}}$. The solution of equation (S5) is simple

$$\tilde{c}(\mathbf{q}, t) = \exp\{-D(1 - \gamma p)\mathbf{q}^2 t\} \exp\left(-\gamma \frac{D}{\lambda^2} t\right). \tag{S6}$$

This expression for the concentration of free excitons in Fourier space will be used later to obtain an analytical expression for the diffusion coefficient.

It is now possible to obtain an exciton's mean-square-displacement MSD$(t)$ = $\sigma^2(t)$ from the second moment of the time dependent probability distribution for the exciton location $C_{tot}(r,t)$ in equation (S4). Without loss of generality, we assume that the average position of the excitons is centered at zero $r$ = 0 (i.e. $\int r\, C_{tot}(r,t)dr^2 = 0$) so that, by definition,

$$\frac{d\sigma^2(t)}{dt} = \int r^2 [\partial_t C_{tot}(r,t)]dr^2 = D \int r^2 [\nabla^2 c(r,t)]dr^2.$$

Integrating twice by parts and using natural boundary conditions (absorbing boundaries) for $C_{tot}(r,t)$,[14] one gets

$$\frac{d\sigma^2(t)}{dt} = 2D \int c(r,t)dr^2. \tag{S7}$$

The time-dependent diffusion coefficient is just the derivative $D(t) = \frac{1}{2}\frac{d\sigma^2(t)}{dt}$. Thus, the combination of equations (S6) and (S7) leads to

$$\begin{aligned}D(t) &= D \int c(r,t)dr^2 = D \int dr^2 \int \tilde{c}(q,t)e^{iq\cdot r}dq = D \int \tilde{c}(q,t)\left[\int e^{iq\cdot r}dr^2\right]dq \\ &= D \int \delta(q)\tilde{c}(q,t)dq = D\tilde{c}(0,t) = D\exp\left(-\gamma\frac{D}{\lambda^2}t\right)\end{aligned} \tag{S8}$$

where in the last step we have used equation (S6) above. Integrating equation (S8) leads to the mean-square-displacement plotted in Figure 2c of the main text,

$$MSD(t) \equiv \sigma^2(t) = 2\frac{\lambda^2}{\gamma}\left[1 - \exp\left(-\gamma\frac{D}{\lambda^2}t\right)\right].$$

The derived exponential decay of the diffusion coefficient is found to be in good agreement with the experimental results with $\gamma$ = 1 and permits the determination of the free diffusion coefficient D and the trap density $1/\lambda^2$. While the derivation of the MSD was done in two dimensions, we used the MSD in one dimension to match the experimental conditions: $MSD(t) = \frac{1}{2}\left(MSD_x(t) + MSD_y(t)\right)$.

**Calculation of sun-equivalent**

The calculation of the sun-equivalent in Figure 2c was performed with the AM1.5 Global (ASTMG173) standard spectra. We extract that 1 sun contains $4.8 \cdot 10^{16}$ photons/s/cm² with an energy larger than the bandgap of (PEA)$_2$PbI$_4$ ($\lambda_{photon} < \lambda_{bandgap} \approx 520\, nm$). Assuming that the absorption of a photon with above-bandgap energy is wavelength independent and that every absorbed photon creates an exciton, one finds a 1 sun equivalent of 25 mW/cm² for a 385 nm light source. In other words, in a first approximation, 1 sun should excite an equal number of excitons in (PEA)$_2$PbI$_4$ as the illumination with 25 mW/cm² at 385 nm.

## 3. Trap state dynamics

**Influence of laser fluence:** Figure S3 shows the lifetime traces of (PEA)$_2$PbI$_4$ recorded with a 10 MHz repetition rate and for different laser fluences of a near-diffraction limited spot. All traces show a multiexponential decay and the same early time dynamics, which shows that exciton-exciton annihilation is absent for laser fluences used in this study. On the other hand, the slow decaying component becomes slightly (logarithmic y-axis) more prominent with increasing laser fluence, which could be due to trap state filling. This is further supported by Figure S3b, where the total emission intensity (integrated lifetime data) is shown as a function of laser fluence. Exciton-exciton annihilation would result in sublinear behavior, but instead a slightly superlinear behavior is observed, which could originate from trap state filling.

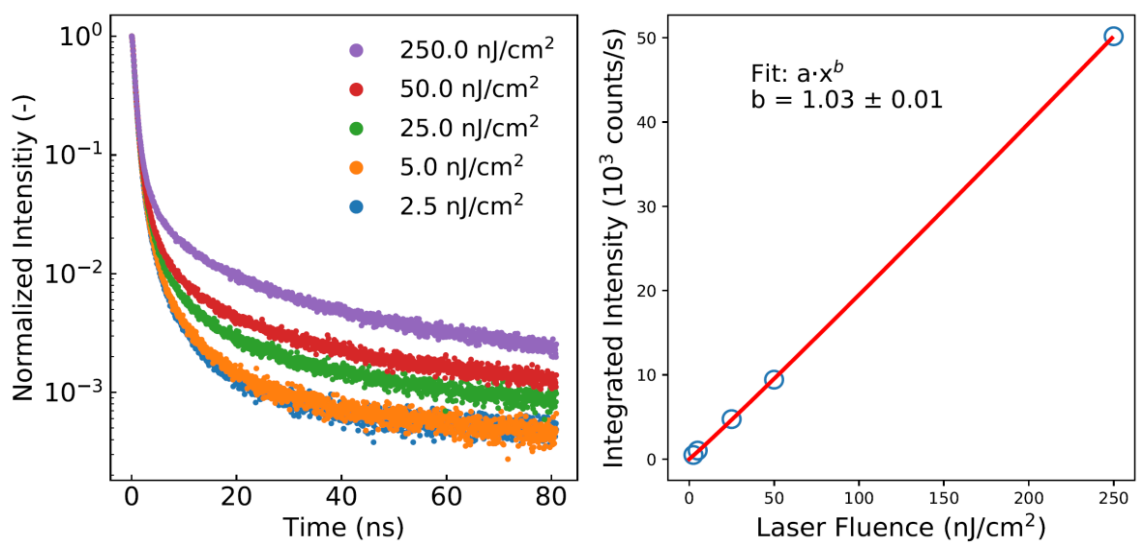

**Figure S3.** (a) Fluorescence lifetime traces of (PEA)$_2$PbI$_4$ for different laser fluences. (b) Fluorescence intensity (integrated lifetime traces from a)) as a function of laser fluence.

**Influence of repetition rate:** The influence or trap state filling is absent in the diffusion measurements performed with laser repetition rates of 5 MHz as shown in Figure S4. For both 50 nJ/cm$^2$ and 250 nJ/cm$^2$, the diffusivity and diffusion exponent are the same.

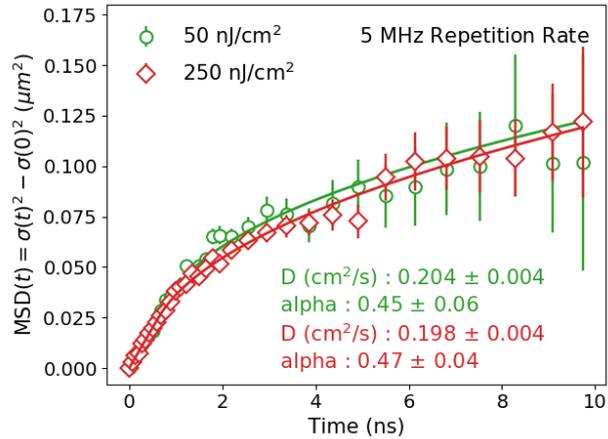

**Figure S4.** MSD(t) from diffusion measurements with a repetition rate of 5 MHz and a laser fluence of 50 (open circles) and 250 nJ/cm$^2$ (open diamonds), respectively.

However, for diffusion measurements performed with a repetition rate of 40 MHz (standard repetition rate of this report) the diffusion exponent increases with increasing laser fluences, while the diffusivity stays constant. An increased diffusion exponent indicates a lower effective trap density, which is likely due to trap state filling. As a result, the intrinsic diffusivity D can be extracted for any laser repetition rate, while it is necessary to use low repetition rates to measure the real trap state density of the perovskites.

The finding that the laser fluence only influences the diffusion measurements for high repetition rates (40 MHz, Figure S5), but not for low ones (5 MHz, Figure S4), suggests that the traps are filled with excitons from previous pulses rather than from excitons which are generated with the same laser pulse. This can be explained by a long lifetime of trap states. With increasing laser fluence a higher percentage of laser pulses create excitons in the same inorganic layer as the following laser pulse, allowing the first exciton to fill a trap and thereby reducing the effective trap density experience by a second exciton generated by a later laser pulse.

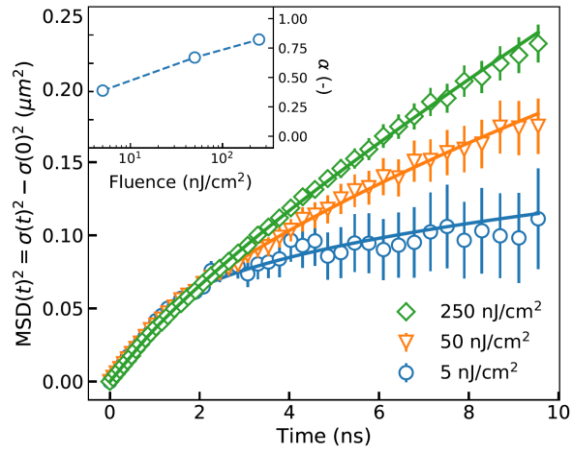

**Figure S5.** Diffusion measurement for different laser fluences with a 40 MHz repetition rate. Higher laser fluences lead to an increased diffusion exponent α (inset), while the intrinsic diffusivity D stays unaffected (slope at early times).

For a better understanding of the trap state filling, we performed diffusion measurements with different repetition rates and a constant fluence of 50 nJ/cm² (Figure S6). All measurements show the same diffusivity and comparable, but slightly increasing diffusion exponents $\alpha$ (Figure S6b). It is possible to reconstruct the standard lifetime trace of the whole laser spot, which does not show any diffusion dynamics, by summing up all the lifetime traces of a diffusion experiment (traces at all positions). However, before the summation, the individual lifetime traces (from different scanning positions) need to be weighted with a factor of $2\pi r$ to account for the whole 2D emission spot, with r being the distance of the APD from the center of the laser spot (see inset in Figure S6c). The resulting lifetime traces are shown in Figure S6c. The lifetime traces acquired with a repetition rate of 2.5, 5, and 10 MHz are almost identical, suggesting a trap state lifetime on the order of 100 ns (= 1/10 MHz).

To guarantee a good signal to noise ratio most measurements in this study were performed with a 40 MHz laser repetition rate. As outlined above, this only changes the observed effective trap density (and thereby diffusion exponent $\alpha$) but does not influence the intrinsic diffusivity of the material.

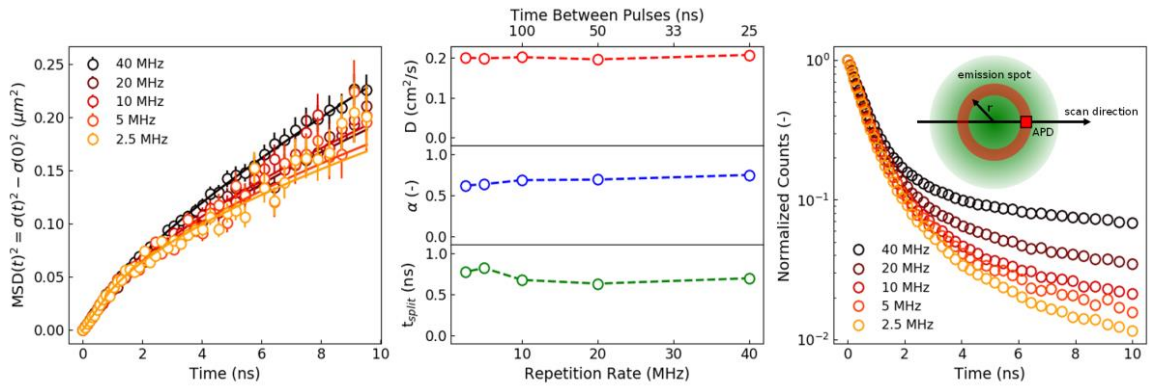

**Figure S6.** (a) MSD(t) from diffusion measurements with a 50 nJ/cm$^2$ fluence and different laser repetition rates. (b) Fit parameters from a): Diffusivity D, diffusion exponent $\alpha$, and split of normal to subdiffusive regime $t_{split}$. (c) Reconstructed lifetime trace from the diffusion data. The inset shows how the different lifetime traces of the diffusion measurement were weighted before summing them up to reconstruct the total fluorescence emission from the laser spot.

## 4. Determination of diffusion length.

Fluorescence lifetime measurements were performed using a laser diode of $\lambda$ = 405 nm (PicoQuant LDH-D-C-405, PDL 800-D, Pico-Harp 300) and an avalanche photodiode (APD, Micro Photon Devices PDM). The repetition rate was 10 MHz and the peak fluence per pulse was 50 nJ/cm$^2$. Figure S7 shows the photoluminescence lifetime traces of (PEA)$_2$PbI$_4$ and (BA)$_2$PbI$_4$, and a tri-exponential fit to the data. The fit was used together with the experimentally obtained time-dependent MSD to extract the total number of surviving excitons for a given time t as presented in Figure 3b of the main text. The total number of surviving excitons at time t is given by:

$$\text{surviving excitons (t)} = \frac{\int_0^t w_1 e^{-\frac{t}{\tau_1}} + w_2 e^{-\frac{t}{\tau_2}} + w_3 e^{-\frac{t}{\tau_3}} \, dt}{\int_0^\infty w_1 e^{-\frac{t}{\tau_1}} + w_2 e^{-\frac{t}{\tau_2}} + w_3 e^{-\frac{t}{\tau_3}} \, dt}$$

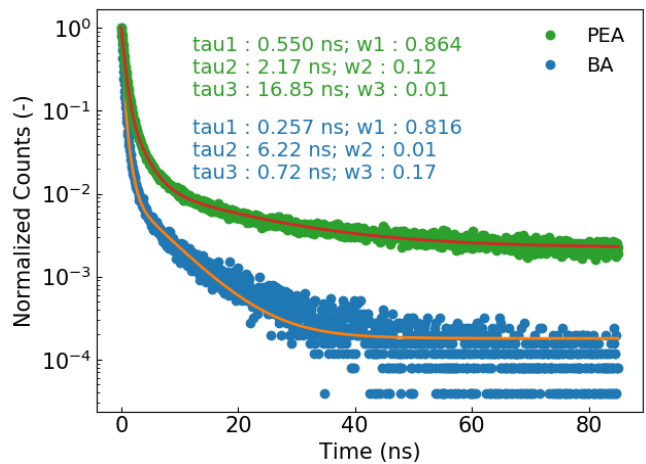

**Figure S7.** Photoluminescence lifetime traces of (PEA)$_2$PbI$_4$ and (BA)$_2$PbI$_4$ with a tri-exponential fit. Fitting parameters are given in the figure.

## 5. Temperature dependent photoluminescence linewidth.

Fluorescence spectra were measured using a spectrograph with a 300 g/mm grating with a blaze of 500 nm (SpectraPro HRS-300) and an EMCCD camera (ProEM HS 1024BX3) from Princeton Instruments. The perovskites were excited with a blue LED (Thorlabs M385PLP1-C5, $\lambda$ = 385 nm). The temperature of the perovskite crystal was varied with a Peltier element (Adaptive Thermal Management, ET-127-10-13-H1), using a PID temperature controller (Dwyer Instruments, Series 16C-3) connected to a type K thermocouple (Labfacility, Z2-K-1M) for feedback control and a fan for cooling. Figure S8 shows the temperature dependent emission spectrum for $(BA)_2PbI_4$ and $(PEA)_2PbI_4$. We applied the Jacobian conversion described by Mooney and Kambhampati to switch from wavelengths to energies.[15]

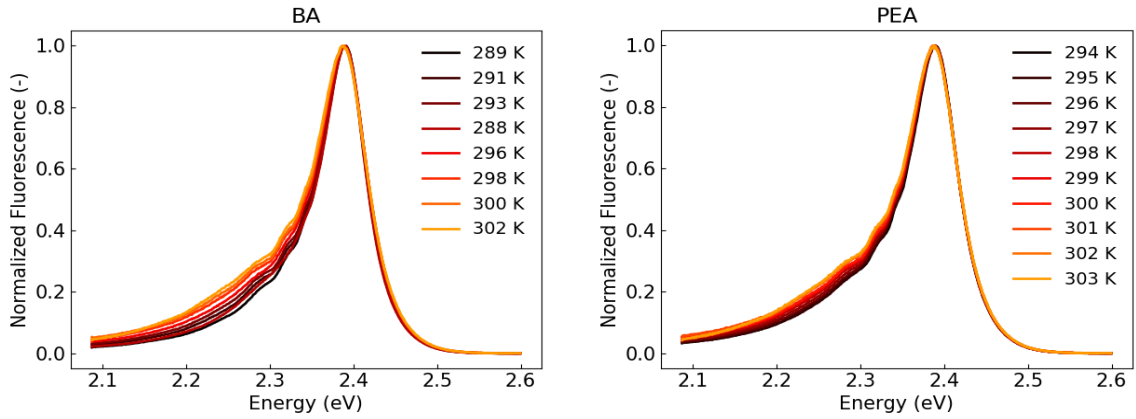

**Figure S8.** Photoluminescence spectra of $(BA)_2PbI_4$ and $(PEA)_2PbI_4$ for different temperatures.

As several studies have shown before, the temperature dependent FWHM $\Gamma(T)$ of photoluminescence of bulk[16–19] and 2D[20–25] perovskites can be described with the Bose−Einstein distribution, due to thermal occupation of phonon modes:

$$\Gamma(T) = \frac{\Gamma_{LO}}{e^{\frac{E_{LO}}{k_B T}} - 1} + \Gamma_0 \quad (S9)$$

where $\Gamma_{LO}$ is the optical phonon coupling strength, $E_{LO}$ is the optical phonon energy, $k_B$ is the Boltzmann constant, and $\Gamma_0$ is the zero phonon linewidth.[26–28] For high temperatures, where the thermal energy is much greater than the phonon energy equation (S9) can be approximated through its asymptote:

$$\Gamma(T) = \frac{\Gamma_{LO}}{e^{\frac{E_{LO}}{k_B T}} - 1} + \Gamma_0 \xrightarrow{k_B T \gg E_{LO}} \frac{\Gamma_{LO}}{E_{LO}} k_B T - \frac{\Gamma_{LO}}{2} + \Gamma_0 \quad (S10)$$

Literature values for $(PEA)_2PbI_4$ and $(BA)_2PbI_4$ range between around 10 to 20 meV for $\Gamma_0$ and 10 to 40 meV for $E_{LO}$. In Figure S9, we plot equation (S9) and its asymptote, equation (S10), for $\Gamma_0 = 0$ and a phonon energy $E_{LO} = 30 meV$. One can see that equation (S9) follows the linear behavior of its asymptote (equation (S10)) already for $T \gtrsim 300K$. As a result, we used equation (S10) to fit our temperature dependent data and extract the optical phonon coupling strength $\Gamma_{LO}$ and optical phonon energy $E_{LO}$ for (PEA)$_2$PbI$_4$ and (BA)$_2$PbI$_4$ (see Figure S10). The resulting $\Gamma_{LO}$ and $E_{LO}$ values are listed in Table S2. $\Gamma_0$ was assumed to be 15 meV in accordance with literature values[20–25,29]. The exact value of $\Gamma_0$ is not critical for the fit, because $\Gamma_{LO}$ is normally significantly greater than $\Gamma_0$. Our values, fit well with the previously reported values for (PEA)$_2$PbI$_4$[20–23,29] and (BA)$_2$PbI$_4$.[24,25] However, we would like to note that due to the uncertainties in this technique the values listed in Table S2 should not be taken as quantitative results, but rather qualitative results to demonstrate the stronger exciton-phonon interaction in (BA)$_2$PbI$_4$ as compared to (PEA)$_2$PbI$_4$.

**Table S2:** Optical phonon coupling strength $\Gamma_{LO}$ and optical phonon energy $E_{LO}$ extracted from temperature dependent photoluminescence data of (PEA)$_2$PbI$_4$ and (BA)$_2$PbI$_4$.

|  | $\Gamma_{LO}\ (meV)$ | $E_{LO}\ (meV)$ |
|---|---|---|
| (PEA)$_2$PbI$_4$ | 126 | 41 |
| (BA)$_2$PbI$_4$ | 532 | 30 |

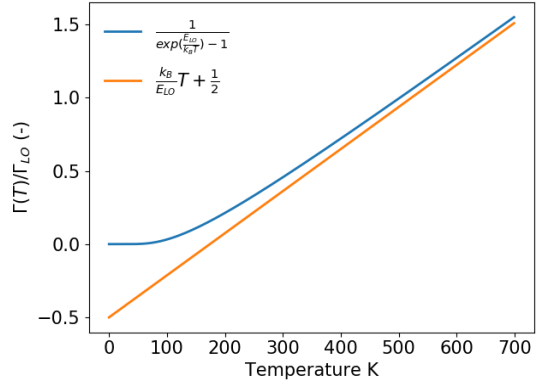

**Figure S9.** Comparison of equation (S9) and its asymptote equation (S10) for $\Gamma_0 = 0$ and $E_{LO} = 30$ meV.

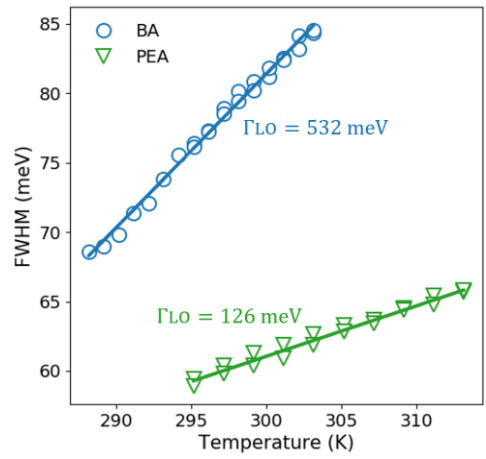

**Figure S10.** Temperature dependent full-width-half-max $\Gamma(T)$ of the photoluminescence of $(PEA)_2PbI_4$ and $(BA)_2PbI_4$. and fits with equation (S10).

## 6. Diffusivities for different chemical compositions

Figure S11 shows the diffusivity values for n = 1 perovskites ($L_2PbI_4$) with different organic spacers L. Plotting the diffusivity values as a function of average atomic displacement $U_{eq}$ reveals an inverse relation and highlights the correlation of lattice softness and diffusivity. Figure S11 shows the plots for average atomic displacements (average $U_{eq} = \frac{1}{\#\,atoms}\sum_i U_{eq}^i$) of the whole perovskite $L_2PbI_4$ (panel b), as well as separated for the organic spacer molecule L (panel c), and the inorganic layer $PbI_4$ (panel d). $U_{eq}$ values were extracted from literature from single crystal x-ray diffraction data: 4-fluoro-phenethylammonium (4FPEA),[6] phenethylammonium (PEA),[4] butylammonium (BA),[5] hexylammonium (HA),[5] octylammonium (OA),[7] decylammonium (DA).[7]

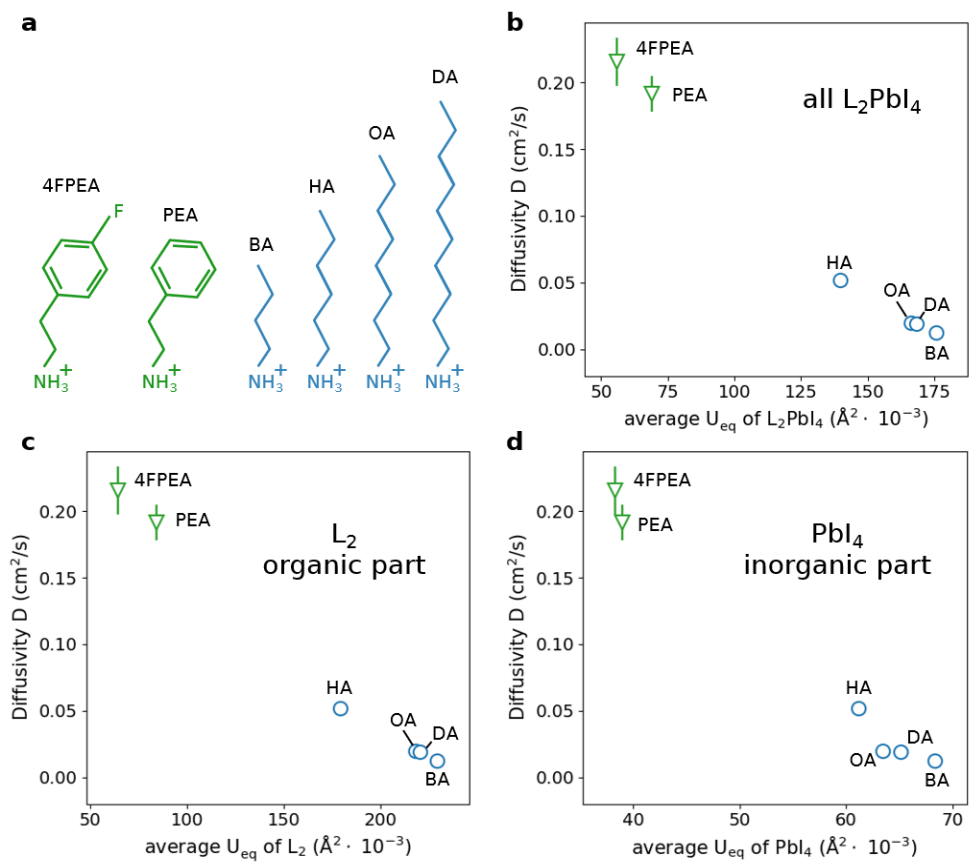

**Figure S11.** (a) Different organic spacers L used in this study (from left to right): 4-fluoro-phenethylammonium (4FPEA), phenethylammonium (PEA), butylammonium (BA), hexylammonium (HA), octylammonium (OA), decylammonium (DA). (b) Diffusivity D vs. average atomic displacement $U_{eq}$ of all the atoms in $L_2PbI_4$ perovskite (average $U_{eq} = \frac{1}{\#\,atoms}\sum_{i=atom}^{L_2PbI_4} U_{eq}^i$). (c). D vs. average $U_{eq}$ of the inorganic part L (average $U_{eq} = \frac{1}{\#\,atoms}\sum_{i=atom}^{L_2} U_{eq}^i$). (d) D vs. average $U_{eq}$ of the inorganic layer $PbI_4$ (average $U_{eq} = \frac{1}{5}U_{eq}^{Pb} + \frac{4}{5}U_{eq}^{I}$).

Figure S12a shows how the diffusivity changes for n = 2 perovskites (L$_2$[APbI$_3$]$_{n-1}$PbI$_4$) for different organic spacers L (PEA and BA) and cations A (methylammonium (MA) and formamidinium (FA)). We find that the organic spacer PEA yields higher diffusivities than BA, just like in the n = 1 case. As for the cation, FA yields higher diffusivities than MA. FA is a larger molecule than MA and fills out the PbI$_6$-octahedra cage more completely. As a result, the PbI$_6$ octahedra are tilted less for FA than for MA. Higher tilt angles were found to yield higher effective electron and hole masses in perovskites.[30,31] In addition, FA being a bulkier molecule results in a more rigid crystal as supported by the average atomic displacement values of the inorganic part (average $U_{eq} = \frac{2}{9}U_{eq}^{Pb} + \frac{7}{9}U_{eq}^{I}$) of (BA)$_2$MAPb$_2$I$_7$ and (BA)$_2$FAPb$_2$I$_7$ being 0.098 and 0.089 Å$^2$, respectively.[8]

It is important to note that Gélvez-Rueda et al. have measured an exciton to free carrier fraction of around 50 % for the n = 2 perovskite (BA)$_2$FAPb$_2$I$_7$, suggesting that transport is not purely excitonic.[32] As a result, our current results serve as a qualitative scaling of the diffusivity and a more rigorous analysis of the n = 2 data would be needed for quantitative diffusivity values.

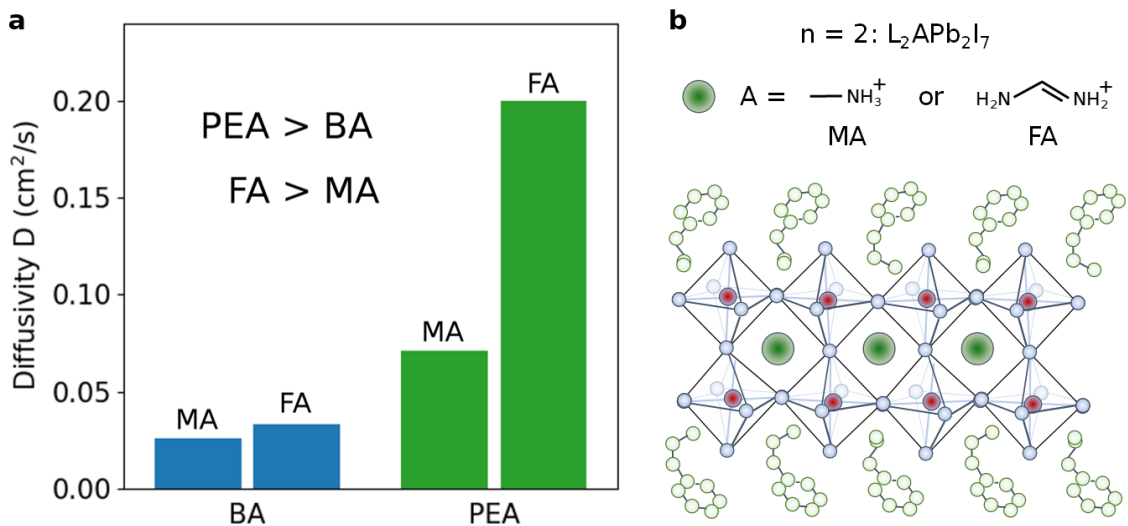

**Figure S12.** (a) Diffusivity for n = 2 perovskites, containing two octahedra per inorganic layer, with chemical formula L$_2$APb$_2$I$_7$. Organic spacers L were PEA or BA. Cation A were methylammonium (MA) or formamidinium (FA). Perovskites with PEA show higher diffusivities than perovskites with BA. Additionally, the diffusivity is higher if FA is used as cation instead of MA. (b) Illustration of a single n = 2 perovskite (L$_2$APb$_2$I$_7$) layer.